\documentclass[
 reprint,
 amsmath,
 amssymb,
 aps,
 superscriptaddress,
 longbibliography,
]{revtex4-2}
\usepackage{graphicx} % Required for inserting images

\usepackage{CJKutf8}
\usepackage[usenames,dvipsnames,table]{xcolor}
\usepackage{graphicx}
\usepackage{amsfonts}
\usepackage{ifsym}
\usepackage{todonotes}
\usepackage{color}
\usepackage{graphicx}
\usepackage{dcolumn}
\usepackage{bm}
\usepackage{hyperref}
\usepackage{booktabs}
\usepackage[normalem]{ulem}
\hypersetup{
    colorlinks=true,       
    linkcolor=blue,          
    citecolor=magenta,        
    filecolor=red,      
    urlcolor=cyan           
}

\usepackage{lmodern}

\usepackage{physics}
\usepackage{amsmath}
\usepackage{xcolor}

\begin{document}

\title{Anomalous Scaling in Shell Model Turbulence }

\author{James Creswell}
\affiliation{
Ludwig Maxmillian University,
Theresienstr. 37, 80333 Munich, Germany
}

\author{Viatcheslav Mukhanov}%
\affiliation{
Ludwig Maxmillian University,
Theresienstr. 37, 80333 Munich, Germany
}

\author{Yaron Oz}%
\affiliation{
School of Physics and Astronomy, Tel-Aviv University, Tel-Aviv 69978, Israel
}
%\affiliation{Simons Center for Geometry and Physics, SUNY, Stony Brook, NY 11794, USA}
%

%
\date{\today}

\begin{abstract}
Shell model turbulence is a simplified mathematical framework that captures essential features of incompressible fluid turbulence
such as the energy cascade, intermittency and anomalous scaling of the fluid observables.
We perform a precision analysis of shell model of a complex velocity field in the steady state turbulent regime, including a calculation of the leading hundred anomalous scaling exponents, 
the probability distribution function of the magnitude and phase of the velocity and the correlations among them
at different shells. We analyze the tail of velocity distribution function  and find that the high moments
exhibit a linear scaling that differs from Kolomogorov's. We explain the origin of this asymptotic scaling 
that offers a new insight to the structure of fluid turbulence.

\end{abstract}

\maketitle
% \tableofcontents

\section{Introduction}

Turbulence is an ubiquitous irregular motion of fluid, such as air or water,
that exhibits rapid and unpredictable changes in velocity, pressure, and density within the fluid. 
Turbulence is inherently difficult to predict and model accurately due to its complex, non-linear nature at a wide range of scales, from large eddies down to very small ones, making it challenging to analyze analytically \cite{Frisch}.
A captivating facet of turbulence is the anomalous scaling of fluid observables at the inertial range of scales, and their deviation from Kolmogorov linear scaling \cite{Kolmogorov1941} due to intermittency.  These exponents hold the key to unraveling the statistical intricacies and structural complexities of turbulent flows.
Analytical derivations of scaling exponents are rare \cite{Falkovich:2009mb}.
A significant challenge arises in accurately measuring these scaling exponents. Despite compelling experimental and numerical evidence showcasing deviations from Kolmogorov scaling, the precision of available data remains insufficient \cite{biferale2019self} to definitively differentiate among the various proposed models in the real world three-dimensional turbulence \cite{PhysRevLett.72.336,PhysRevE.63.026307,Eling:2015mxa,oz2017spontaneous}. 
Hence, the pursuit of precision turbulence emerges as an imperative paradigm shift in turbulence research. 

Shell models provide a simplified mathematical framework to study fluid turbulence
\cite{1973SPhD...18..216G,1989PThPh..81..329O,PhysRevE.58.1811}, and they capture some of the essential features of turbulence, while reducing the complexity of the equations involved. This framework can be particularly useful to model the energy cascade, intermittency and the statistical properties of turbulent fluid flows.
Compared to solving the Navier-Stokes equations, which are computationally expensive, shell models are computationally efficient, making them natural playgrounds for developing precision 
turbulence.
%Of course, one expects that properties such as 
%anomalous scaling have a different structure than the three-dimensional one.
Numerical calculations of scaling exponent in shell models have been carries out for the low moments
of the velocity probability distribution function in
\cite{1995PhFl....7..617K,PhysRevE.58.1811,PhysRevE.62.8037,PhysRevE.68.046304,doi:10.1146/annurev.fluid.35.101101.161122}.
A hidden symmetry in shell models has been proposed in \cite{physRevFluids.7.034604}.

The aim of this letter is to address precision turbulence in the framework of the complex velocity
shell model introduced in \cite{PhysRevE.58.1811}.
We will perform a detailed analysis of shell model of the complex velocity field in the steady state turbulent regime, including a calculation of the leading hundred anomalous scaling exponents, 
the probability distribution function of the velocity field magnitude and its phase and the correlations among the velocities
at different shells. We will analyze the tail of velocity distribution function  and show that the high moments of the distribution exhibit a linear scaling that differs from Kolomogorov's.
We will uncover the origin of this asymptotic linear scaling and offer a  new insight to the structure of fluid turbulence.

The letter is organized as follows.
In section II we briefly review the SABRA shell model for a complex velocity field, present the anomalous scaling exponents $\zeta_p$
up to $p=100$ and discuss the large $p$ asymptotics.
In section III we consider the joint probability distribution function 
of the the velocities at different shells including the magnitude and phase marginal distributions and their correlation structure.
Section IV is devoted to a discussion.
Details of the calculations and additional plots are given in the Supplemental Materials.

\section{Turbulence Scaling in Shell Model}

\subsection{SABRA model}
We will consider the SABRA shell model \cite{PhysRevE.58.1811}.
It
consists of complex valued shells velocity scale field indexed by the shell number, $u_n$, evolving in time according to:
\begin{equation}
    \label{eq:model}
    \begin{split}
    \frac{du_n}{dt} = i \left(a k_{n+1} u_{n+2} u_{n+1}^* + b k_n u_{n+1} u_{n-1}^* \right. \\ 
    \left. - c k_{n-1} u_{n-1} u_{n-2} \right) - \nu k_n^2 u_n + f_n \ ,
    \end{split}
\end{equation}
where $n = 1, 2, \dots$ indexes the shells.
$a$, $b$, and $c$ are real-valued constants, $k_n$ are wavenumbers obeying $k_n = k_0 \lambda^n$ for some constants $k_0$ and $\lambda$, and $\nu$ is a viscosity parameter.
$f_n$ describes an external forcing, taken to be Gaussian random noise, which will be applied to the first shells (IR). Energy conservation dictates that $a+b+c=0$.
The shell model reaches a steady state and exhibits anomalous scaling at the inertial range of scale $n_f\ll n \ll n_{\nu}$ 
where $n_f$ and $n_{\nu}$ correspond to the forcing scale in the IR and the viscous scale in the UV.
One searches for a scaling
\begin{equation}
    \label{eq:model}
    \expval{|u_n|^p} \propto k_n^{-\zeta_p} \ ,
\end{equation}
where the average is taken over the space of solutions in the steady state turbulent regime.
The scaling exponents $\zeta_p$ are believed to be universal and independent of the force and viscous structures.
Our main  goal is to determine accurately these scaling exponents up to $p=100$, using simulations in which equation \eqref{eq:model} is solved numerically.

% \section{Technicalities}

Our choice of parameters and forcing follows \cite{PhysRevE.58.1811}:
$a = 1,  b = c =  -0.5, \lambda =2, k_0 = 2^{-4},  \nu = 4\times 10^{-11}$.
We will have a total of 50 shells, although not all will be part of the data analysis of the inertial range. 
% Once the solution data are obtained, the scaling exponents $\zeta_q$ are determined using the fitting procedure in L'vov et al., equations (44) and (45) in that paper.
% However, I observe that a simple power law fit of $\expval{|u_n|^q}$ in the range $n = 3$ to $n = 24$ works nearly as well.
The forcing term $f_n$ is applied to the first two shells (see details in the Supplementary Materials).
Although the details of the forcing are not expected to change the results we are seeking, to be explicit we are using the same type of coloured (correlated) Gaussian random noise described in \cite{PhysRevE.58.1811}.
For the initial condition,  the energy is divided  between the first two shells randomly:
\begin{align}
    u_1 = \sqrt{\alpha E_0 } e^{2 \pi i \beta},
    u_2 = \sqrt{(1 - \alpha) E_0 } e^{2 \pi i \gamma} \ ,
\end{align}
An initial energy $E_0 = 10$ is chosen, and $\alpha$, $\beta$, $\gamma$ are drawn uniformly from the unit interval.
%$f_n$ is partly correlated in time.
% \begin{equation}
%     \label{eq:model}
%     \begin{split}
%     f_n(t + dt) = f_n(t) e^{-dt/\tau} \\ + \sigma_n \sqrt{-2 (1 - e^{-2 dt / \tau}) \log_{10}(\alpha)} e^{2 \pi i \beta}.
%     \end{split}
% \end{equation}
%$dt$ is the timestep, $\tau$ is the forcing time-scale, $\tau \sim 1 /(k_n u_n)$, $\alpha$ and $\beta$ are two %uniform random numbers between $0$ and $1$ generated at each step, and
%$\sigma_1 = 5\times10^{-3}$, $\sigma_2 = \sigma_1 \sqrt{-c/a}$.
%(The choice of $\sigma_2 / \sigma_1 \approx 0.7$ is supposed to reduce the helicity flux.)
%\YO{Need to explain that we checked other parameters and other forcings}
For completeness, we check the robustness of the calculated scaling exponents for different values of $E_0$ and $\nu$, as well
as different procedures for generating the forcing including correlated noise and white noise.
Further information is available in the Supplementary Materials.

\subsection{Scaling Exponents}
 A universal structure of turbulence is expected only at the inertial range of scales $n_f\ll n \ll n_{\nu}$. 
In order to determine this range we use the analytical result for the third moment $\zeta_3=1$
which serves as a reference point to evaluate the numerical the simulations.
The departure of $\zeta_3$ from $1$ at a given accuracy gives the breakdown of the inertial range to that accuracy. This is shown in the Supplementary Materials.

In Table I we list the moments scaling exponents up to hundred and we plot them in Fig. \ref{fig:enter-label_p}.
The low moments agree with the results of \cite{PhysRevE.58.1811}.
The fit to Eq.~(\ref{eq:model}) is performed using least-squares regression. 
The errors are calculated as
$$ \mathrm{std}(\zeta_p) = \frac{\text{best-fit residual}}{\text{degrees of freedom}} \times C,$$
where $C$ is the component of the covariance matrix corresponding to the parameter $\zeta_p$.
This quantifies the statistical error of the fit. 
% There is an another error, associated with different random fluctuations in the simulation data.
% This error, which is partly due to the finite data size, depends on the total length of data averaged in the exponent calculation. 
% By calculating the exponents in different subsections of the data separately, we can estimate this error, and see when it stabilizes.

\begin{table}[h!]
    \centering
    % \footnotesize
    \begin{tabular}{|c|c|}
    \hline
    $p$ & $\zeta_p$  \\
    \hline
    \hline
    1 & $0.402 \pm 0.004$\\
    2 & $0.730 \pm 0.008$\\
    3 & $1.002 \pm 0.012$\\
    4 & $1.272 \pm 0.019$\\
    5 & $1.513 \pm 0.031$\\
    6 & $1.744 \pm 0.048$\\
    7 & $1.968 \pm 0.065$\\
    8 & $2.188 \pm 0.082$\\
    9 & $2.407 \pm 0.098$\\
    10 & $2.624 \pm 0.114$\\
    11 & $2.841 \pm 0.129$\\
    12 & $3.057 \pm 0.144$\\
    13 & $3.273 \pm 0.159$\\
    14 & $3.490 \pm 0.178$\\
    15 & $3.706 \pm 0.196$\\
    16 & $3.923 \pm 0.214$\\
    17 & $4.140 \pm 0.232$\\
    18 & $4.356 \pm 0.250$\\
    19 & $4.573 \pm 0.269$\\
    20 & $4.790 \pm 0.286$\\
    21 & $5.007 \pm 0.304$\\
    22 & $5.225 \pm 0.322$\\
     \hline 
    \end{tabular}
    \begin{tabular}{|c|c|}

    \hline
    $p$ & $\zeta_p$  \\
    \hline
    \hline
    23 & $5.442 \pm 0.340$\\
    24 & $5.660 \pm 0.358$\\
    25 & $5.877 \pm 0.375$\\
    26 & $6.095 \pm 0.393$\\
    27 & $6.313 \pm 0.410$\\
    28 & $6.530 \pm 0.428$\\
    29 & $6.748 \pm 0.446$\\
    30 & $6.966 \pm 0.463$\\
    35 & $7.757 \pm 0.551$\\
    40 & $8.838 \pm 0.638$\\
    45 & $9.921 \pm 0.724$\\
    50 & $11.004 \pm 0.811$\\
    55 & $12.088 \pm 0.897$\\
    60 & $13.172 \pm 0.983$\\
    65 & $14.256 \pm 1.069$\\
    70 & $15.341 \pm 1.154$\\
    75 & $16.426 \pm 1.235$\\
    80 & $17.511 \pm 1.317$\\
    85 & $18.597 \pm 1.399$\\
    90 & $19.683 \pm 1.481$\\
    95 & $20.768 \pm 1.562$\\
    100 & $21.854 \pm 1.644$\\ \hline 
    \end{tabular}
\end{table}

%The estimation is based on fitting to the spectrum as discussed in [1].

% \begin{figure}[h!]
%     \centering
%     \includegraphics[width=0.75\textwidth]{means2.pdf}
%     \caption{Same as Fig. 3, up to $q = 16$.}
%     \label{fig:my_label}
% \end{figure}

\ \\

%\subsubsection{High Moments}
%We have calculated exponents and corresponding uncertainties up to $p=50$.

%\subsection{Fit for scaling exponents}

\begin{figure}
    \centering% \includegraphics[width=0.5\textwidth]{fig1.pdf}
    \includegraphics[width=0.5\textwidth]{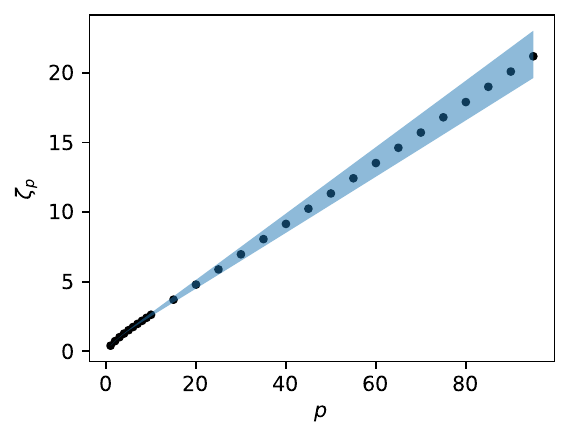}
    \caption{Scaling exponents up to $p=100$, with the 90\% confidence region shown in blue. The uncertainty is the systematic error (see the appendix for discussion).}
    \label{fig:enter-label_p}
\end{figure}

\subsection{Large $p$ Asymptotics}

\begin{figure}[h!]
    \centering
    \includegraphics[width=0.48\textwidth]{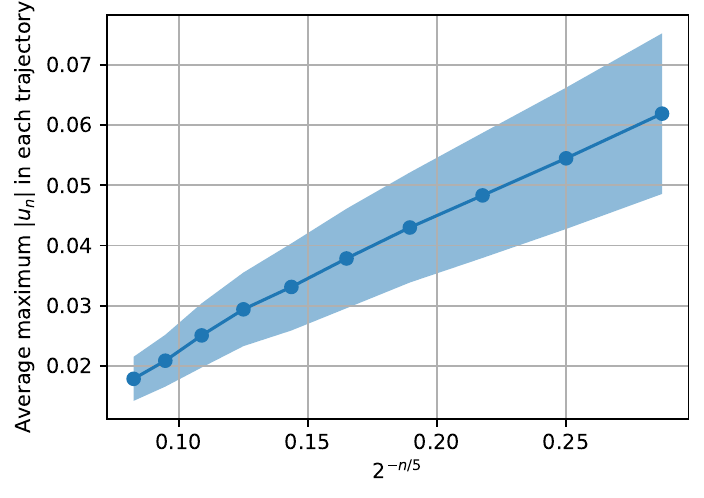}
    \caption{Maximal velocity magnitude $|u_n|$ in each shell versus $k_n^{-\frac{1}{5}} = 2^{-n/5}$  where $n$ is the shell number. We see a linear scaling.}
    \label{fig:maxs-shells}
\end{figure}

Fig. \ref{fig:enter-label_p} indicates that the large $p$ limit of the anomalous scaling exponents
is $\zeta_p = \frac{p}{5}$, which is a linear scaling that differs from Kolmogorov's  $\zeta_p = \frac{p}{3}$.
In order to understand the origin of this asymptotic scaling consider the 
marginal probability distribution functions (PDFs) of $|u_n|$, $f(|u_n|)$. The main contribution to the higher moments comes
from the maximal value $|u_n| = |u_{max}|$. More precisely, the PDFs $|u_n|^p f(|u_n|)$ at large $p$  have a sharp peak at $|u_n|\sim \alpha k_n^{-\frac{1}{5}} = \alpha 2^{-\frac{n}{5}}$, where $\alpha$ is a constant which is independent of $p$, see Fig. \ref{fig:enter-label_p}, and
Fig. \ref{fig:maxs-shells} where we plot the maximal $|u_n|$ in each shell.

\begin{figure}[h!]
    \centering
    \includegraphics[width=0.4\textwidth]{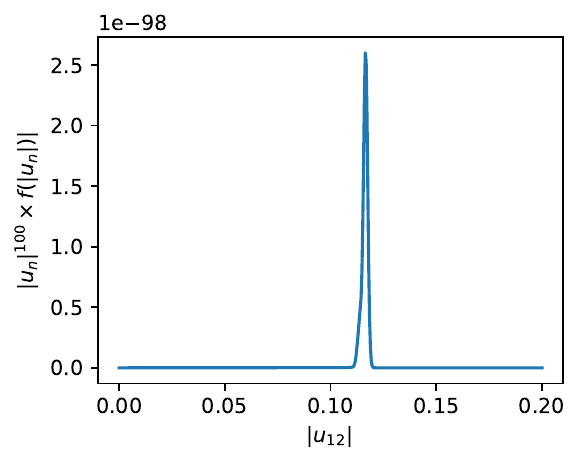}
    \caption{Moment distribution function, $|u_n|^p \times f(|u_n|)$, for $p=100$, where $f(|u_n|)$ is the PDF of the velocity magnitude. The 12th shell ($n=12$) is plotted.}
    \label{fig:enter-label_p}
\end{figure}

This reveals and interesting structure of turbulence scaling and leads 
us to propose that
\begin{equation}
\lim_{p\to\infty} \left(|u_n|^p f(|u_n|)\right)  = \delta(|u_n|-\alpha k_n^{-\frac{1}{5}}) \  ,
\end{equation}
where $\alpha$ is a non-universal constant.

%\newpage 

%\newpage 

\section{Velocity Distribution Function}

%In this section we carry out further statistical analysis to characterize the simulated data.
%We can examine the distribution of the $u_n$ data in two ways: the variations of $u_n$ between shells and, within each %shell, the distribution of the time-ordered data.

%\subsection{Energy Distribution in Shells}
%Firstly we consider the relative size and behaviour between different shells.
% The normalized distribution function of the log-normal distribution is:
% \begin{equation}
%     f(x) = \frac{1}{x \sigma \sqrt{2 \pi}} \exp(-\frac{(\log(x) - \mu)^2}{2 \sigma^2}). 
% \end{equation}
% We try to fit the function $f(n') = k_{n'}^2 |u_{n'}|^2$ to the log-normal distribution, with $n' = n_\mathrm{max} - n$ as the independent variable.
% $f(n')$ is normalized, and the fit is only carried out for $n$ in the ``inertial'' range.
% The best fit parameters are $\mu = 2.03$, $\sigma = 0.18$.
% The fit is not perfect, and the parameters could be affected by noise and hyperparameter decisions.
% See Figure 7.

%In this section we consider properties of the shell velocities joint probability distribution function.

%\subsection{\boldmath Marginal Probability Distribution Functions} 

In the Supplemental Materials  we plot the marginal probability distributions functions of the velocity magnitude $|u_n|$ and its phase.
%, which are the histograms over all time series values $t > 500$ at fixed $n$.
The magnitude PDFs can be approximated by a log-normal distribution for small velocities
with the skewness increasing with $n$. However, the log-normal distribution does not capture 
correctly the fast drop of the distribution at higher velocity magnitudes and its tail. 
%This is expected since, being a time independent observable at steady state, we do not expect the PDFs %to be local functions.
%See more details in the supplemental material.
The marginal phase PDFs are uniform, as expected \cite{PhysRevE.58.1811}.

\subsection{Covariance matrix}

\begin{figure}
    \centering
    \includegraphics[width=0.48\textwidth]{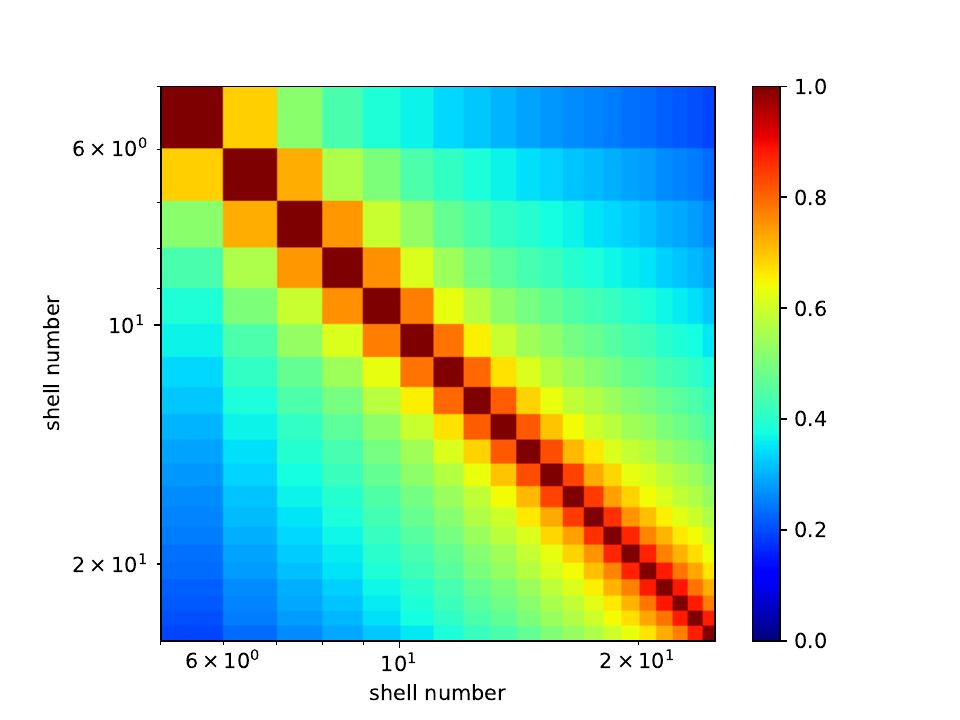}
    \includegraphics[width=0.48\textwidth]{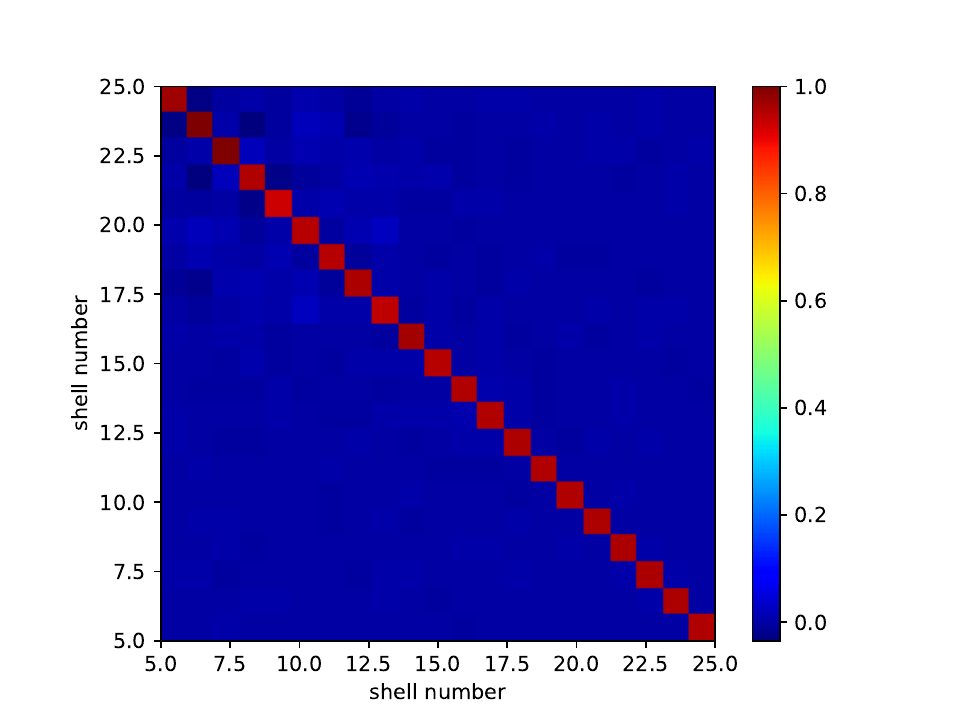}
    \caption{Equal-time covariance of the velocity magnitude (upper plot) and phase (lower plot) between shells in the inertial range.
    %In the upper panel the axes are displayed logarithmically to correspond to a linear physical size %scale.
    }
    \label{fig:cov-matrix}
\end{figure}

%The most general correlation analysis is the estimation of the full covariance matrix itself.
%As pointed out in [1], the phases in the model are constrained by a fixed relation therefore the correlation functions %of the phases or quantities involving them should be 0. (page 4)
%This we verified with our new empirical data.
%However, the cross-correlation of the amplitudes, $\expval{|u_n| |u_m}$, is not necessarily 0.
The covariance matrices of velocity magnitudes and phases in different shells are shown in Fig. \ref{fig:cov-matrix}.
In both cases the covariance is the normalized equal-time correlation function, i.e., the Pearson cross-correlation coefficient:
\begin{equation}
    C(n, m) = \frac{\mathrm{cov}(n, m)}{\sigma_n \sigma_m} \ ,
\end{equation}
where cov is the covariance estimated by multiplying equal times of two shells $u_n$ and $u_m$, and $\sigma_n$ and $\sigma_m$ are the corresponding variances of each shell.
The values of $C$ range from $-1$ to $1$, where $1$ means perfectly correlated.
As seen in Fig. \ref{fig:cov-matrix}, there is a spreading of the off-diagonal velocity magnitude covariance $C(n, m)$ between two shells $n$ and $m$ as $n$ and $m$ increase.
%This reflects the logarithmic arrangement of the shell sizes.
Putting the axes in logarithmic scales, the covariance matrix takes a uniform structure. The phase corvariance matrix is diagonal, hence the phases at different shells are independent.
%This is shown in Figure 14.
%, identical except for a possible positive rescaling).

\section{Discussion}

We performed a precision analysis of shell model of a complex velocity field in the steady state turbulent regime and calculated the leading hundred anomalous scaling exponents, 
the probability distribution function of the velocity magnitude and its phase and the correlations among the velocities
at different shells. We found that the high moments of the velocity magnitude PDF  
exhibit a linear scaling that differs from Kolomogorov's, whose origin is 
the dominant contribution from the maximum velocity.
This provides an interesting new insight on the rare events of the turbulence in this model.
Phenomenologically we found that a formula of the type \cite{PhysRevLett.72.336}
\begin{equation} \label{eq:fp}
    f(p) = \frac{p}{5} + \frac{1}{2} \qty(1 - (0.2)^{p/3}) \ ,
\end{equation}
fits nicely the numerical data.

%In an upcoming work we will present similar analysis for the real-space shell model (equation (4)).

\vspace{0.2cm}
{\bf Note added} While typing the manuscript we received 
\cite{dewit2024extreme}, which contains some overlap with our calculation of the scaling exponents.

\section*{Acknowledgements}
This work is supported in part by the Israeli Science Foundation Excellence Center, the US-Israel Binational Science Foundation, the Israel Ministry of Science and the LMU-TAU International Research Grant.

\appendix
%\onecolumngrid
Further details and plots concerning the numerical analysis, correlation analysis, and other technical details are in the attached Supplementary Materials.

\bibliography{arxiv_main.bib}

%apsrev4-2.bst 2019-01-14 (MD) hand-edited version of apsrev4-1.bst
%Control: key (0)
%Control: author (8) initials jnrlst
%Control: editor formatted (1) identically to author
%Control: production of article title (0) allowed
%Control: page (0) single
%Control: year (1) truncated
%Control: production of eprint (0) enabled
\begin{thebibliography}{17}%
\makeatletter
\providecommand \@ifxundefined [1]{%
 \@ifx{#1\undefined}
}%
\providecommand \@ifnum [1]{%
 \ifnum #1\expandafter \@firstoftwo
 \else \expandafter \@secondoftwo
 \fi
}%
\providecommand \@ifx [1]{%
 \ifx #1\expandafter \@firstoftwo
 \else \expandafter \@secondoftwo
 \fi
}%
\providecommand \natexlab [1]{#1}%
\providecommand \enquote  [1]{``#1''}%
\providecommand \bibnamefont  [1]{#1}%
\providecommand \bibfnamefont [1]{#1}%
\providecommand \citenamefont [1]{#1}%
\providecommand \href@noop [0]{\@secondoftwo}%
\providecommand \href [0]{\begingroup \@sanitize@url \@href}%
\providecommand \@href[1]{\@@startlink{#1}\@@href}%
\providecommand \@@href[1]{\endgroup#1\@@endlink}%
\providecommand \@sanitize@url [0]{\catcode `\\12\catcode `\$12\catcode
  `\&12\catcode `\#12\catcode `\^12\catcode `\_12\catcode `\%12\relax}%
\providecommand \@@startlink[1]{}%
\providecommand \@@endlink[0]{}%
\providecommand \url  [0]{\begingroup\@sanitize@url \@url }%
\providecommand \@url [1]{\endgroup\@href {#1}{\urlprefix }}%
\providecommand \urlprefix  [0]{URL }%
\providecommand \Eprint [0]{\href }%
\providecommand \doibase [0]{https://doi.org/}%
\providecommand \selectlanguage [0]{\@gobble}%
\providecommand \bibinfo  [0]{\@secondoftwo}%
\providecommand \bibfield  [0]{\@secondoftwo}%
\providecommand \translation [1]{[#1]}%
\providecommand \BibitemOpen [0]{}%
\providecommand \bibitemStop [0]{}%
\providecommand \bibitemNoStop [0]{.\EOS\space}%
\providecommand \EOS [0]{\spacefactor3000\relax}%
\providecommand \BibitemShut  [1]{\csname bibitem#1\endcsname}%
\let\auto@bib@innerbib\@empty
%</preamble>
\bibitem [{\citenamefont {Frisch}\ and\ \citenamefont
  {Kolmogorov}(1995)}]{Frisch}%
  \BibitemOpen
  \bibfield  {author} {\bibinfo {author} {\bibfnamefont {U.}~\bibnamefont
  {Frisch}}\ and\ \bibinfo {author} {\bibfnamefont {A.~N.}\ \bibnamefont
  {Kolmogorov}},\ }\href@noop {} {\emph {\bibinfo {title} {Turbulence: the
  legacy of AN Kolmogorov}}}\ (\bibinfo  {publisher} {Cambridge university
  press},\ \bibinfo {year} {1995})\BibitemShut {NoStop}%
\bibitem [{\citenamefont {{Kolmogorov}}(1941)}]{Kolmogorov1941}%
  \BibitemOpen
  \bibfield  {author} {\bibinfo {author} {\bibfnamefont {A.}~\bibnamefont
  {{Kolmogorov}}},\ }\bibfield  {title} {\bibinfo {title} {{The Local Structure
  of Turbulence in Incompressible Viscous Fluid for Very Large Reynolds'
  Numbers}},\ }\href@noop {} {\bibfield  {journal} {\bibinfo  {journal}
  {Akademiia Nauk SSSR Doklady}\ }\textbf {\bibinfo {volume} {30}},\ \bibinfo
  {pages} {301} (\bibinfo {year} {1941})}\BibitemShut {NoStop}%
\bibitem [{\citenamefont {Falkovich}\ \emph {et~al.}(2010)\citenamefont
  {Falkovich}, \citenamefont {Fouxon},\ and\ \citenamefont
  {Oz}}]{Falkovich:2009mb}%
  \BibitemOpen
  \bibfield  {author} {\bibinfo {author} {\bibfnamefont {G.}~\bibnamefont
  {Falkovich}}, \bibinfo {author} {\bibfnamefont {I.}~\bibnamefont {Fouxon}},\
  and\ \bibinfo {author} {\bibfnamefont {Y.}~\bibnamefont {Oz}},\ }\bibfield
  {title} {\bibinfo {title} {{New relations for correlation functions in
  Navier-Stokes turbulence}},\ }\href
  {https://doi.org/10.1017/S0022112009993429} {\bibfield  {journal} {\bibinfo
  {journal} {J. Fluid Mech.}\ }\textbf {\bibinfo {volume} {644}},\ \bibinfo
  {pages} {465} (\bibinfo {year} {2010})},\ \Eprint
  {https://arxiv.org/abs/0909.3404} {arXiv:0909.3404 [nlin.CD]} \BibitemShut
  {NoStop}%
\bibitem [{\citenamefont {Biferale}\ \emph {et~al.}(2019)\citenamefont
  {Biferale}, \citenamefont {Bonaccorso}, \citenamefont {Buzzicotti},\ and\
  \citenamefont {Iyer}}]{biferale2019self}%
  \BibitemOpen
  \bibfield  {author} {\bibinfo {author} {\bibfnamefont {L.}~\bibnamefont
  {Biferale}}, \bibinfo {author} {\bibfnamefont {F.}~\bibnamefont
  {Bonaccorso}}, \bibinfo {author} {\bibfnamefont {M.}~\bibnamefont
  {Buzzicotti}},\ and\ \bibinfo {author} {\bibfnamefont {K.~P.}\ \bibnamefont
  {Iyer}},\ }\bibfield  {title} {\bibinfo {title} {Self-similar subgrid-scale
  models for inertial range turbulence and accurate measurements of
  intermittency},\ }\href@noop {} {\bibfield  {journal} {\bibinfo  {journal}
  {Physical review letters}\ }\textbf {\bibinfo {volume} {123}},\ \bibinfo
  {pages} {014503} (\bibinfo {year} {2019})}\BibitemShut {NoStop}%
\bibitem [{\citenamefont {She}\ and\ \citenamefont
  {Leveque}(1994)}]{PhysRevLett.72.336}%
  \BibitemOpen
  \bibfield  {author} {\bibinfo {author} {\bibfnamefont {Z.-S.}\ \bibnamefont
  {She}}\ and\ \bibinfo {author} {\bibfnamefont {E.}~\bibnamefont {Leveque}},\
  }\bibfield  {title} {\bibinfo {title} {Universal scaling laws in fully
  developed turbulence},\ }\href {https://doi.org/10.1103/PhysRevLett.72.336}
  {\bibfield  {journal} {\bibinfo  {journal} {Phys. Rev. Lett.}\ }\textbf
  {\bibinfo {volume} {72}},\ \bibinfo {pages} {336} (\bibinfo {year}
  {1994})}\BibitemShut {NoStop}%
\bibitem [{\citenamefont {Yakhot}(2001)}]{PhysRevE.63.026307}%
  \BibitemOpen
  \bibfield  {author} {\bibinfo {author} {\bibfnamefont {V.}~\bibnamefont
  {Yakhot}},\ }\bibfield  {title} {\bibinfo {title} {Mean-field approximation
  and a small parameter in turbulence theory},\ }\href
  {https://doi.org/10.1103/PhysRevE.63.026307} {\bibfield  {journal} {\bibinfo
  {journal} {Phys. Rev. E}\ }\textbf {\bibinfo {volume} {63}},\ \bibinfo
  {pages} {026307} (\bibinfo {year} {2001})}\BibitemShut {NoStop}%
\bibitem [{\citenamefont {Eling}\ and\ \citenamefont
  {Oz}(2015)}]{Eling:2015mxa}%
  \BibitemOpen
  \bibfield  {author} {\bibinfo {author} {\bibfnamefont {C.}~\bibnamefont
  {Eling}}\ and\ \bibinfo {author} {\bibfnamefont {Y.}~\bibnamefont {Oz}},\
  }\bibfield  {title} {\bibinfo {title} {{The Anomalous Scaling Exponents of
  Turbulence in General Dimension from Random Geometry}},\ }\href
  {https://doi.org/10.1007/JHEP09(2015)150} {\bibfield  {journal} {\bibinfo
  {journal} {JHEP}\ }\textbf {\bibinfo {volume} {09}},\ \bibinfo {pages}
  {150}},\ \Eprint {https://arxiv.org/abs/1502.03069} {arXiv:1502.03069
  [nlin.CD]} \BibitemShut {NoStop}%
\bibitem [{\citenamefont {Oz}(2017)}]{oz2017spontaneous}%
  \BibitemOpen
  \bibfield  {author} {\bibinfo {author} {\bibfnamefont {Y.}~\bibnamefont
  {Oz}},\ }\bibfield  {title} {\bibinfo {title} {Spontaneous symmetry breaking,
  conformal anomaly and incompressible fluid turbulence},\ }\href@noop {}
  {\bibfield  {journal} {\bibinfo  {journal} {Journal of High Energy Physics}\
  }\textbf {\bibinfo {volume} {2017}},\ \bibinfo {pages} {1} (\bibinfo {year}
  {2017})}\BibitemShut {NoStop}%
\bibitem [{\citenamefont {{Gledzer}}(1973)}]{1973SPhD...18..216G}%
  \BibitemOpen
  \bibfield  {author} {\bibinfo {author} {\bibfnamefont {E.~B.}\ \bibnamefont
  {{Gledzer}}},\ }\bibfield  {title} {\bibinfo {title} {{System of hydrodynamic
  type admitting two quadratic integrals of motion}},\ }\href@noop {}
  {\bibfield  {journal} {\bibinfo  {journal} {Soviet Physics Doklady}\ }\textbf
  {\bibinfo {volume} {18}},\ \bibinfo {pages} {216} (\bibinfo {year}
  {1973})}\BibitemShut {NoStop}%
\bibitem [{\citenamefont {{Ohkitani}}\ and\ \citenamefont
  {{Yamada}}(1989)}]{1989PThPh..81..329O}%
  \BibitemOpen
  \bibfield  {author} {\bibinfo {author} {\bibfnamefont {K.}~\bibnamefont
  {{Ohkitani}}}\ and\ \bibinfo {author} {\bibfnamefont {M.}~\bibnamefont
  {{Yamada}}},\ }\bibfield  {title} {\bibinfo {title} {{Temporal Intermittency
  in the Energy Cascade Process and Local Lyapunov Analysis in Fully-Developed
  Model Turbulence}},\ }\href {https://doi.org/10.1143/PTP.81.329} {\bibfield
  {journal} {\bibinfo  {journal} {Progress of Theoretical Physics}\ }\textbf
  {\bibinfo {volume} {81}},\ \bibinfo {pages} {329} (\bibinfo {year}
  {1989})}\BibitemShut {NoStop}%
\bibitem [{\citenamefont {L'vov}\ \emph {et~al.}(1998)\citenamefont {L'vov},
  \citenamefont {Podivilov}, \citenamefont {Pomyalov}, \citenamefont
  {Procaccia},\ and\ \citenamefont {Vandembroucq}}]{PhysRevE.58.1811}%
  \BibitemOpen
  \bibfield  {author} {\bibinfo {author} {\bibfnamefont {V.~S.}\ \bibnamefont
  {L'vov}}, \bibinfo {author} {\bibfnamefont {E.}~\bibnamefont {Podivilov}},
  \bibinfo {author} {\bibfnamefont {A.}~\bibnamefont {Pomyalov}}, \bibinfo
  {author} {\bibfnamefont {I.}~\bibnamefont {Procaccia}},\ and\ \bibinfo
  {author} {\bibfnamefont {D.}~\bibnamefont {Vandembroucq}},\ }\bibfield
  {title} {\bibinfo {title} {Improved shell model of turbulence},\ }\href
  {https://doi.org/10.1103/PhysRevE.58.1811} {\bibfield  {journal} {\bibinfo
  {journal} {Phys. Rev. E}\ }\textbf {\bibinfo {volume} {58}},\ \bibinfo
  {pages} {1811} (\bibinfo {year} {1998})}\BibitemShut {NoStop}%
\bibitem [{\citenamefont {{Kadanoff}}\ \emph {et~al.}(1995)\citenamefont
  {{Kadanoff}}, \citenamefont {{Lohse}}, \citenamefont {{Wang}},\ and\
  \citenamefont {{Benzi}}}]{1995PhFl....7..617K}%
  \BibitemOpen
  \bibfield  {author} {\bibinfo {author} {\bibfnamefont {L.}~\bibnamefont
  {{Kadanoff}}}, \bibinfo {author} {\bibfnamefont {D.}~\bibnamefont {{Lohse}}},
  \bibinfo {author} {\bibfnamefont {J.}~\bibnamefont {{Wang}}},\ and\ \bibinfo
  {author} {\bibfnamefont {R.}~\bibnamefont {{Benzi}}},\ }\bibfield  {title}
  {\bibinfo {title} {{Scaling and dissipation in the GOY shell model}},\ }\href
  {https://doi.org/10.1063/1.868775} {\bibfield  {journal} {\bibinfo  {journal}
  {Physics of Fluids}\ }\textbf {\bibinfo {volume} {7}},\ \bibinfo {pages}
  {617} (\bibinfo {year} {1995})},\ \Eprint
  {https://arxiv.org/abs/chao-dyn/9409001} {arXiv:chao-dyn/9409001 [nlin.CD]}
  \BibitemShut {NoStop}%
\bibitem [{\citenamefont {L'vov}\ and\ \citenamefont
  {Procaccia}(2000)}]{PhysRevE.62.8037}%
  \BibitemOpen
  \bibfield  {author} {\bibinfo {author} {\bibfnamefont {V.~S.}\ \bibnamefont
  {L'vov}}\ and\ \bibinfo {author} {\bibfnamefont {I.}~\bibnamefont
  {Procaccia}},\ }\bibfield  {title} {\bibinfo {title} {Analytic calculation of
  the anomalous exponents in turbulence: Using the fusion rules to flush out a
  small parameter},\ }\href {https://doi.org/10.1103/PhysRevE.62.8037}
  {\bibfield  {journal} {\bibinfo  {journal} {Phys. Rev. E}\ }\textbf {\bibinfo
  {volume} {62}},\ \bibinfo {pages} {8037} (\bibinfo {year}
  {2000})}\BibitemShut {NoStop}%
\bibitem [{\citenamefont {Benzi}\ \emph {et~al.}(2003)\citenamefont {Benzi},
  \citenamefont {Biferale}, \citenamefont {Sbragaglia},\ and\ \citenamefont
  {Toschi}}]{PhysRevE.68.046304}%
  \BibitemOpen
  \bibfield  {author} {\bibinfo {author} {\bibfnamefont {R.}~\bibnamefont
  {Benzi}}, \bibinfo {author} {\bibfnamefont {L.}~\bibnamefont {Biferale}},
  \bibinfo {author} {\bibfnamefont {M.}~\bibnamefont {Sbragaglia}},\ and\
  \bibinfo {author} {\bibfnamefont {F.}~\bibnamefont {Toschi}},\ }\bibfield
  {title} {\bibinfo {title} {Intermittency in turbulence: Computing the scaling
  exponents in shell models},\ }\href
  {https://doi.org/10.1103/PhysRevE.68.046304} {\bibfield  {journal} {\bibinfo
  {journal} {Phys. Rev. E}\ }\textbf {\bibinfo {volume} {68}},\ \bibinfo
  {pages} {046304} (\bibinfo {year} {2003})}\BibitemShut {NoStop}%
\bibitem [{\citenamefont
  {Biferale}(2003)}]{doi:10.1146/annurev.fluid.35.101101.161122}%
  \BibitemOpen
  \bibfield  {author} {\bibinfo {author} {\bibfnamefont {L.}~\bibnamefont
  {Biferale}},\ }\bibfield  {title} {\bibinfo {title} {Shell models of energy
  cascade in turbulence},\ }\href
  {https://doi.org/10.1146/annurev.fluid.35.101101.161122} {\bibfield
  {journal} {\bibinfo  {journal} {Annual Review of Fluid Mechanics}\ }\textbf
  {\bibinfo {volume} {35}},\ \bibinfo {pages} {441} (\bibinfo {year}
  {2003})}\BibitemShut {NoStop}%
\bibitem [{\citenamefont {Mailybaev}(2022)}]{physRevFluids.7.034604}%
  \BibitemOpen
  \bibfield  {author} {\bibinfo {author} {\bibfnamefont {A.~A.}\ \bibnamefont
  {Mailybaev}},\ }\bibfield  {title} {\bibinfo {title} {Shell model
  intermittency is the hidden self-similarity},\ }\href
  {https://doi.org/10.1103/PhysRevFluids.7.034604} {\bibfield  {journal}
  {\bibinfo  {journal} {Phys. Rev. Fluids}\ }\textbf {\bibinfo {volume} {7}},\
  \bibinfo {pages} {034604} (\bibinfo {year} {2022})}\BibitemShut {NoStop}%
\bibitem [{\citenamefont {de~Wit}\ \emph {et~al.}(2024)\citenamefont {de~Wit},
  \citenamefont {Ortali}, \citenamefont {Corbetta}, \citenamefont {Mailybaev},
  \citenamefont {Biferale},\ and\ \citenamefont {Toschi}}]{dewit2024extreme}%
  \BibitemOpen
  \bibfield  {author} {\bibinfo {author} {\bibfnamefont {X.~M.}\ \bibnamefont
  {de~Wit}}, \bibinfo {author} {\bibfnamefont {G.}~\bibnamefont {Ortali}},
  \bibinfo {author} {\bibfnamefont {A.}~\bibnamefont {Corbetta}}, \bibinfo
  {author} {\bibfnamefont {A.~A.}\ \bibnamefont {Mailybaev}}, \bibinfo {author}
  {\bibfnamefont {L.}~\bibnamefont {Biferale}},\ and\ \bibinfo {author}
  {\bibfnamefont {F.}~\bibnamefont {Toschi}},\ }\href@noop {} {\bibinfo {title}
  {Extreme statistics and extreme events in dynamical models of turbulence}}
  (\bibinfo {year} {2024}),\ \Eprint {https://arxiv.org/abs/2402.02994}
  {arXiv:2402.02994 [physics.flu-dyn]} \BibitemShut {NoStop}%
\end{thebibliography}%

\newpage 
\ \\
\newpage
\appendix 
\section{Supplementary Materials}
\subsection{Simulation}
In this section we describe the procedure for generating the simulated data.
The equation (1),
\begin{eqnarray}
    \label{eq:model}
    \frac{du_n}{dt} = i (a k_{n+1} u_{n+2} u_{n+1}^{*} + b k_n u_{n+1} u_{n-1}^{*} \nonumber\\
    - c k_{n-1} u_{n-1} u_{n-2} ) - \nu k_n^2 u_n + f_n ,
\end{eqnarray}
where $n = 1, 2, \dots$ indexes the shells,
for the given parameters in use as described in the paper.
Because the scale of $u_n$ varies in different shells over multiple orders of magnitude, the system of differential equations becomes very stiff, i.e.\ numerical methods of integration are prone to instability and inaccuracy even with small step sizes.
Implicit integration methods, in which previously computed data points are used to estimate the derivative at the current computation, improve the stability and robustness against stiffness. In particular, we choose a backward differentiation formula (BDF) method, with an adaptive order and adaptive time-step size, to perform the integration.

\subsection{Noise}
$f_n$ describes an external forcing, taken to be Gaussian random noise, which is only applied to the first shells.
Although the details of the forcing are not expected to change the results we are seeking, to be explicit we are using the same type of coloured (correlated) Gaussian random noise described in [1].
For the initial condition, a starting energy $E_0 = 10$ is chosen and divided between the first two shells randomly:
\begin{eqnarray}
    u_1(t=0) &=& \sqrt{\alpha E_0 } e^{2 \pi i \beta}, \nonumber\\
    u_2(t=0) &=& \sqrt{(1 - \alpha) E_0 } e^{2 \pi i \gamma} \ ,
\end{eqnarray}
where $\alpha$, $\beta$, $\gamma$ are drawn uniformly from the unit interval.
For the time evolution, $f_n$ obeys:
\begin{eqnarray}
&f_n(t + dt) = f_n(t) e^{-dt/\tau} +  \nonumber\\
&+ \sigma_n \sqrt{-2 (1 - e^{-2 dt / \tau})\log_{10}(\alpha)} e^{2 \pi i \beta} \ , 
\end{eqnarray}
where $dt$ is the timestep, $\tau$ is the forcing time-scale, $\tau \sim 1 /(k_n u_n)$, and $\alpha$ and $\beta$ are two uniform random numbers between $0$ and $1$ generated at each step.
Following [1], we used $\sigma_1 = 5\times10^{-3}$ and $\sigma_2 = \sigma_1 \sqrt{-c/a}. $

In addition to the procedure for generating noise and initial conditions described above and used in the main simulations and analysis in the paper, we have also tried several different choices for the forcing and other parameters of the setup.
This includes adjusting the viscosity $\nu$, adjusting the initial energy $E_0$, and different types of noise, including uncorrelated white noise ($\tau \rightarrow \infty$).
In no case did these variations change the scaling exponents or other overall statistics. 
However, they did affect the time required to reach steady state.

\subsection{Error analysis}
There are two errors which apply to the calculation of the scaling exponents from finite data:
\begin{enumerate}
    \item The uncertainty in the best-fit parameter of the slope.
    \item The variation in the slope in finite segments of the data.
\end{enumerate}
The first kind of error is being estimated using the standard least-squares error.
The second kind of error is estimated using a procedure, in which the dataset is further subdivided into sections, and the fit performed on the partial datasets, which is quoted in Table I in the main paper.

% \subsection{Parameters}

\section{Determination of the Inertial Range}
\label{inertial}
\begin{figure*}[h!]
    \centering
    \includegraphics[width=0.45\textwidth]{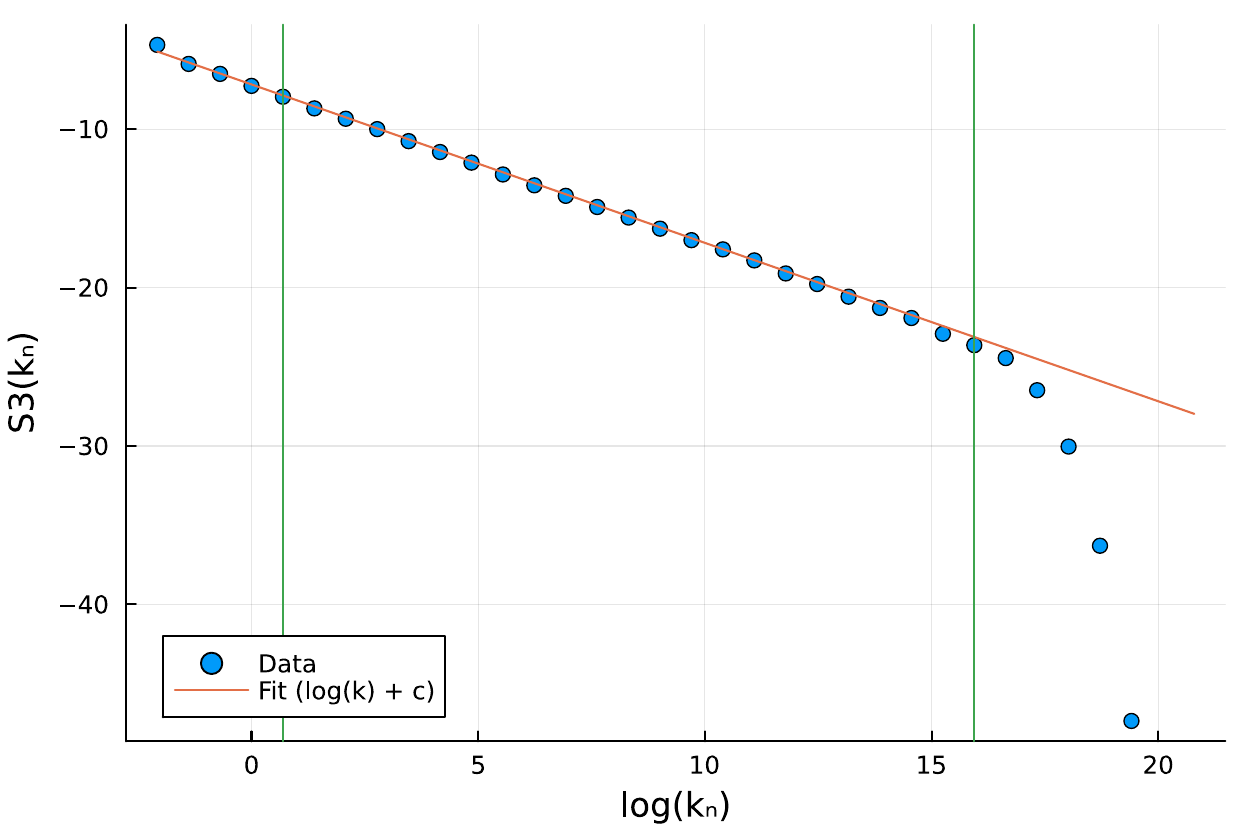}
    \includegraphics[width=0.45\textwidth]{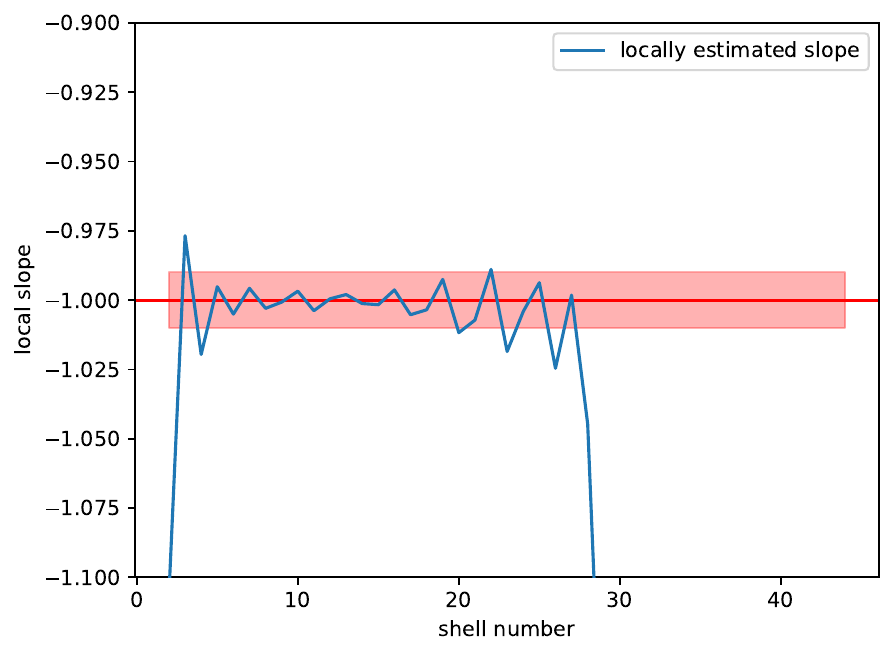}
    \caption{Left: $S_3 = \expval{|u_n|^3}$ spectrum. Right: Determination of the inertial range: $\zeta_3$ within 1\% of the analytically known result for $p=3$.}
    \label{fig:inertial_range}
\end{figure*}

The scaling power law is only expected within the  
inertial range of scale $n_f\ll n \ll n_{\nu}$ 
where $n_f$ and $n_{\nu}$ correspond to the forcing scale in the IR and the viscous scale in the UV.
For $p = 3$, an exact scaling $\zeta_p = 1$ can be derived analytically and this is
used to determine the inertial range.
The departure from $\zeta_p = 1$ at a given accuracy gives the breakdown of the inertial range to that accuracy.
In Fig. \ref{fig:inertial_range} the spectrum for $p=3$ and the range is zoomed around $1$ to determine the inertial range.
Adopting 1\% as an accuracy threshold (corresponding to the red shaded region in \ref{fig:inertial_range}), the simulations permit about 25 shells in the inertial range.
In particular, shells 5 through 25 are retained for analysis below.
In order to reach steady state,  the data from $t=0$ to $t=500$ of each simulation is also excluded from the calculations of the scaling exponents in all analysis.

%%%

\section{Large $p$ Asymptotics}
\label{p}
\begin{figure*}
    \centering
    \includegraphics[width=0.98\textwidth]{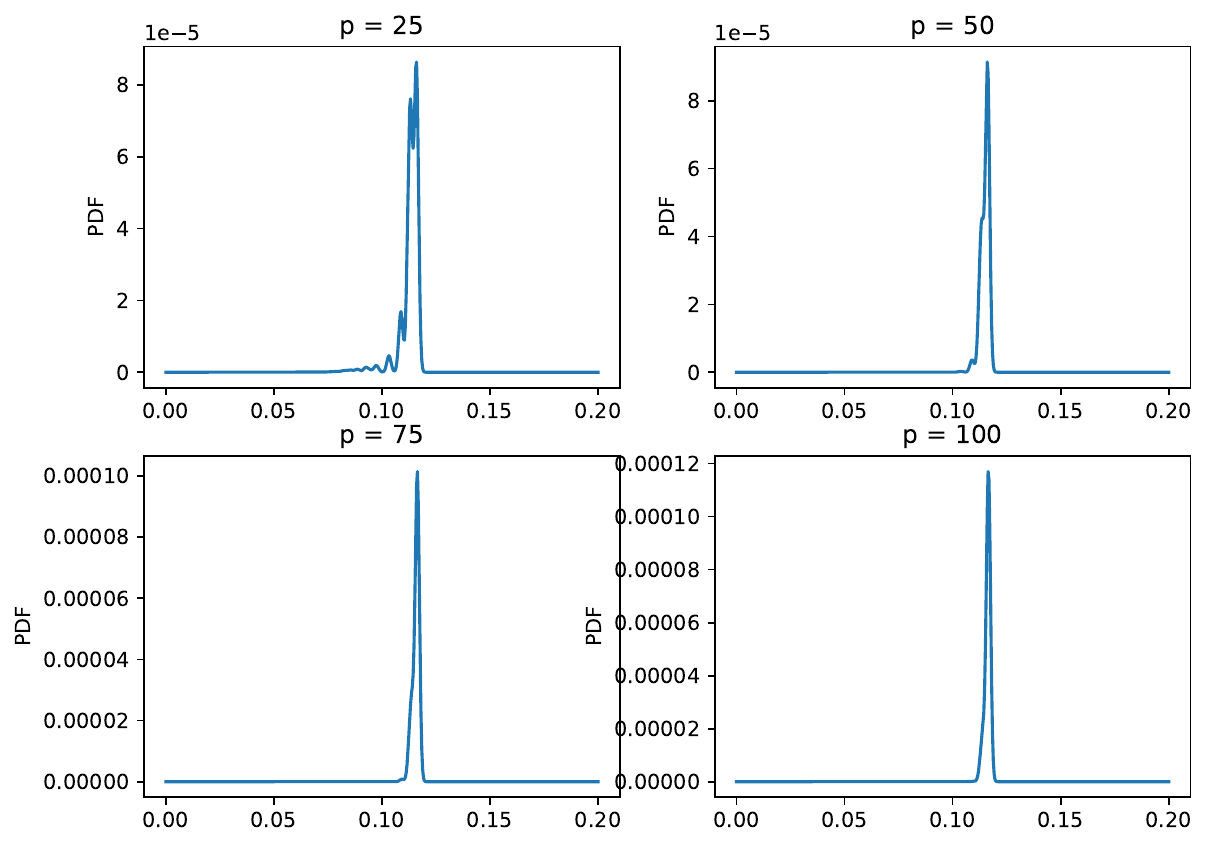}
    \caption{As the power $p$ increases, the $p$th moment integrand, $|u_n|^p \times f(|u_n|)$, where $f$ is the distribution function, approaches to a narrow peak around the maximum velocity. In this figure the trend is shown for the 12th shell. The maximum velocity in this shell is $0.116$.}
    \label{fig:delta}
\end{figure*}

As the power $p$ increases, the $p$th moment integrand, $|u_n|^p \times f(|u_n|)$, where $f$ is the velocity magnitude probability distribution function, approaches to a narrow peak around the maximum velocity.
This is illustrated in Fig. \ref{fig:delta}, where the moment integrand is plotted for $p=25, 50, 75, 100$.

%A consistency check can be applied to verify that the dominant contribution to the scaling exponents $\zeta_p$ in the large-$p$ limit comes from the tail of the velocity distribution, i.e. those parts of the %time ordered data where the velocity is the highest.
%It was seen in Figure 1 that the scaling exponents as a function of $p$ approach a line with slope $p/5$, %hence:
%$$\expval{|u_n|} \propto k_n^{-p/5} = 2^{-np/5}.$$
% At the same time, the maximum $u$ in each shell scales like 

% \begin{align*}
%     \mathrm{max}(|u_n|) &= 2 \times 2^{-n/5} \\
%     \mathrm{max}(|u_n|)^p &= 2^p \times (2^{-n})^{p/5} \\
%     &\stackrel{?}{=} C k_n^{-\zeta_p} = C (2^{-n})^{p/5}
% \end{align*}

%\begin{figure}
%    \centering
%    \includegraphics[width=0.45\textwidth]{max-us-check.pdf}
%    \caption{In the inertial range, the maximum $|u_n|$ scales as it should to explain the observed exponents.}
%    \label{fig:}
%\end{figure}

\section{Marginal Distribution of Velocities Magnitudes and Phases}
\label{marginal}
\begin{figure}[h!]
    \centering
    \includegraphics[width=0.48\textwidth]{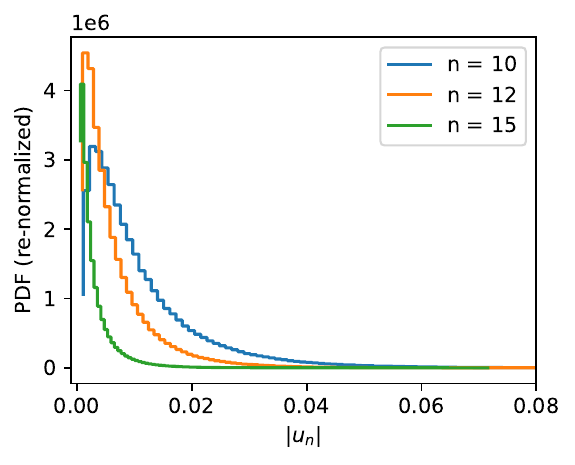}
    \caption{Normalized marginal PDF of the velocity magnitude $|u_n|$ for the shells $n  = 10, 12, 15$.}
    \label{fig:enter-label_mag}
\end{figure}

\begin{figure*}[h!]
    \centering
    \includegraphics[width=\textwidth]{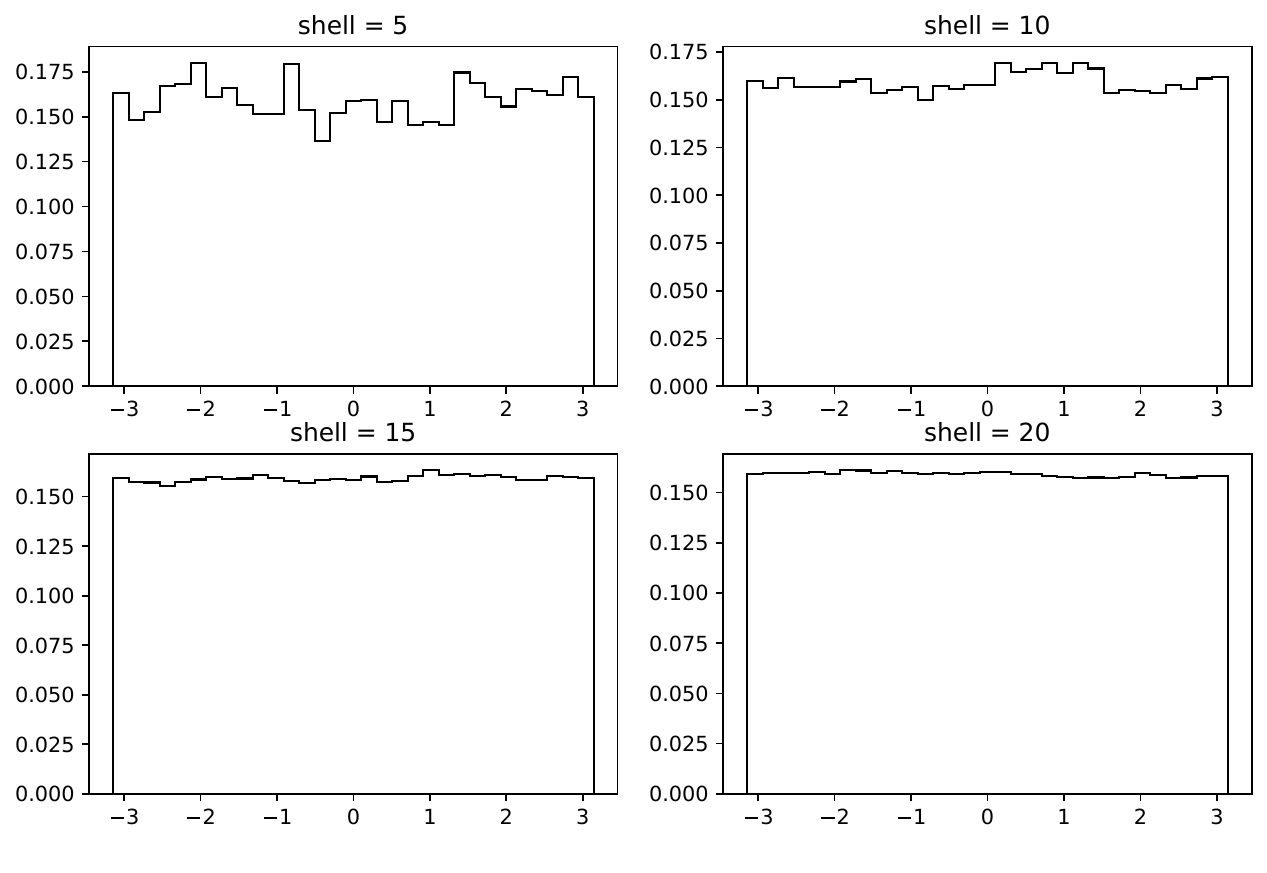}
    \caption{Histograms of phases in each of several shells (marginals). }
    \label{fig:phases}
\end{figure*}

Complementary to the correlation analysis included in the paper, here we show the marginal distribution functions of the magnitude Fig. \ref{fig:enter-label_mag} and 
phases Fig. \ref{fig:phases} of the complex-valued $u_n$ trajectories.

%Figure~\ref{fig:phases}

\section{Local Energy Dissipation}
\begin{figure}[h!]
    \centering
    \includegraphics[width=0.48\textwidth]{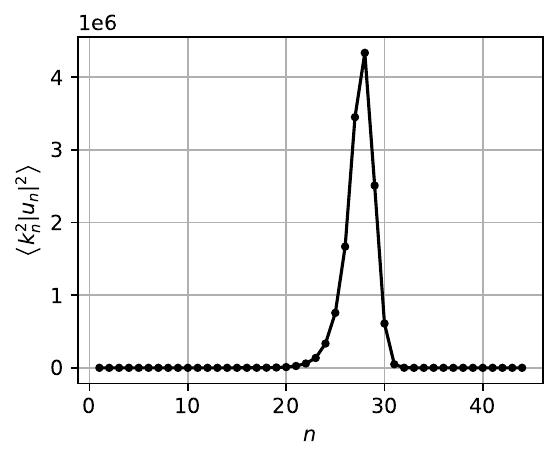}
    \includegraphics[width=0.48\textwidth]{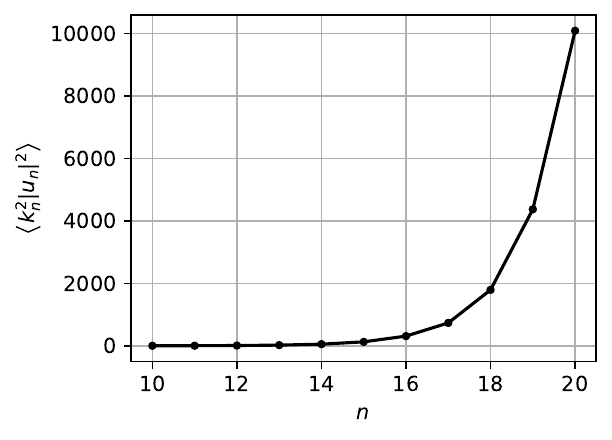}
    \caption{The mean $k_n^2 |u_n|^2$ in each shell. Upper panel: the full range of shells. Lower panel: Zoom only on the shells of the inertial range.}
    \label{fig:my_label_diss}
\end{figure}

In Fig. \ref{fig:my_label_diss} we plot the average local energy dissipation $k_n^2 |u_n|^2$ at each shell. 
We see a pick at the transition shells from the inertial range to the viscous regime.
Restricting to the inertial range we get the expected scaling
$k_n^2 |u_n|^2 \sim k_n^{2-\xi_2}$. 

\end{document}